\documentclass{sf2a-conf2015}
\usepackage{graphicx}
\usepackage{hyperref}
\usepackage[]{natbib}  
\usepackage{epstopdf}

\def\BibTeX{{\rm B\kern-.05em{\sc i\kern-.025em b}\kern-.08em
    T\kern-.1667em\lower.7ex\hbox{E}\kern-.125emX}}
\bibpunct{(}{)}{;}{a}{}{,}  %%%%%%%%%%%%%  A&A bibliography style
\begin{document}

\TitreGlobal{SF2A 2015}

%%-----------------------------------------------------------------
%%      the top matter
%%

\title{Discovery of new Chemically Peculiar late B-type stars: HD 67044}

\runningtitle{HD 67044}

\author{R.Monier$^{1,}$}\address{LESIA, UMR 8109, Observatoire de Paris Meudon, Place J.Janssen, Meudon, France}\address{Lagrange, UMR 7293, Universite de Nice Sophia, Nice, France}
\author{M.Gebran}\address{Department of Physics and Astronomy, Notre Dame University - Louaize, PO Box 72, Zouk Mikael, Lebanon}
\author{F.Royer}\address{GEPI, UMR 8111, Observatoire de Paris Meudon, Place J.Janssen, Meudon, France}
%\author{E.Griffin}\address{Dominion Astrophysical Observatory, 5071 West Saanich Road, Victoria, BC, V9E 2E7, Canada} 

%% IF Author3 has two affiliations, the one of Author1 and a second one:
%\author{C.\,E. Author3$^{1,}$}\address{Dept. of Chess, University of Games, 35101 Las Vegas, Monaco} 

%% Keep this line, even if the page will be settled afterwards.
\setcounter{page}{237}

%%-----------------------------------------------------------------

\maketitle

%%-----------------------------------------------------------------
%%        The abstract
%% 
%%  Warning!  within the abstract:
%%  - do not use macros. 
%%  - do not use commands like: \cite, \citet, \citep ... etc.

\begin{abstract}
Using high dispersion high quality spectra of HD 67044 obtained with the echelle spectrograph SOPHIE at Observatoire de Haute Provence, we show that this star contains strong lines of Silicon, Titanium, Chromium, Yttrium, Zirconium and Europium. Line synthesis of these lines yield overabundances which range from 3 $\odot$ up to 200 $\odot$. We therefore propose that HD 67044 be reclassified as a late Chemically Peculiar B star of the SiCrEu type.
\end{abstract}

%% Insert the keywords (to appear in the ADS indexing)
%% Keywords must be separated by a comma

%%-----------------------------------------------------------------

\section{Introduction}
%%---------------------

HD 67044 currently assigned a B8 spectral type is one of the slowly rotating B stars situated in the northern hemisphere which we are currently observing. The selection criteria for this sample of stars are a declination higher than $-15^{o}$, spectral class B8 or B9, luminosity class V or IV, and a magnitude V brighter than 7.85. Most of the stars of this B8-9 sample have just recently been observed in December 2014. We are currently performing a careful abundance analysis study of high resolution high $\frac{S}{N}$ ratio spectra of these objects and sort them out into chemically normal stars (ie. whose abundances do not depart more than $\pm$ 0.15 dex from solar), new spectroscopic binaries and new chemically peculiar B stars (CPs) which had remained unoticed so far.% HD 67044 turns out to be a new CP star.
We present here new abundance determinations for HD 67044 which allow us to propose that this star is a new CP late B star. 
\cite{Monier15} have recently published the discovery of 4 new HgMn stars (3 from this late-B stars sample and one from a sample of 47 early A types stars verifying the same criteria).
\cite{Royer14} have published the analysis of the sample of 47 early A stars having low apparent projected velocities in the northern hemisphere up to V=6.65 mag.
A careful abundance analysis of high resolution high $\frac{S}{N}$ ratio spectra of these objects has
sorted out the sample into 17 chemically normal stars, 12 spectroscopic binaries and 13 Chemically Peculiar stars (CPs) among which 5 are new CP stars.

\section{Observations and reduction}

HD 67044 has been observed once at Observatoire de Haute Provence using the High Resolution (R =75000) mode of SOPHIE in December 2014. A 40 minutes exposure was secured in December 2014 with a $\frac{S}{N}$ ratio of about 130.

\section{The nature of the new CP star HD 67044}

Several spectral regions have been used to readdress the spectral type of HD 67044. The star being a late B-type dwarf, the chemical peculiarity could be either i) of the HgMn type, ii) or of the Si type or iii) of the SrCrEu type, or iv) a hybrid of the last two.
We therefore investigated several spectral regions containing strong resonance or low excitation lines of Hg, Mn, Si, Sr, Cr and Eu.
First, the red wing of $H_{\epsilon}$ harbors the Hg II $\lambda$ 3984 \AA\ line and several Zr II and Y II lines likely to be strenghtened in late B star of the Hg-Mn type. The region from 4125 \AA\ to 4145 \AA\ contains the classification Si II doublet (M 2), the Mn II line at 4136 \AA\ and the Sr II resonance line at 4129.72 \AA\ likely to be strengthened respectively in a Bp Si star, in a star enriched in Mn, and a SrCrEu Bp star.
The regions 4070-4080 \AA, 4210-4220 \AA\ and 4300-4310 \AA\ contain the resonance lines of Sr II at 4077.71 \AA\ and 4215.52 \AA\ and the low-excitation line at 4305.44 \AA.
The 4200-4210 \AA\ region contains the Eu II resonance line at 4205.04 \AA. The 4550-4560 \AA\ region contains the strongest expected Cr II line at 4558.65 \AA.
We fail to detect the Hg II line at 3984 \AA. The Mn II line at 4136.92 \AA\ is not particularly strong, nor are the Sr II lines. In contrast, the lines of Si II (Multiplet 2), Cr II, Y II , Zr Ii and Eu II are strong in HD 67044. This leads us to rule out that HD 67044 be a new HgMn star. It probably rather is a new CP star of the SiCrEu type.

% Redo the table 1
%------------------------------------------------
\begin{table}
\centering
\begin{tabular}{|l|l|l|p{3cm}|}
 \hline
  \multicolumn{4}{|c|}{\textbf{Classification lines}}   \\ \hline
  \multicolumn{1}{|c|}{Laboratory Wavelengths (\AA)} &
  \multicolumn{1}{|c|}{Identification} &
  \multicolumn{1}{|c|}{Multiplet} &
  \multicolumn{1}{|c|}{Abundance}  \\ \hline
3982.44 & Y II &  & 200 $\odot$\\
 3983.87 & Hg II & M 2 & not detected \\
 3990.96 & Zr II &  &  70 $\odot$\\
3998.82  & Zr II &  &  70 $\odot$ \\
4077.71 & Sr II &   &  1  $\odot$  \\
4128.07 & Si II & M 2 & 2-3 $\odot$\\
%4129.72 & Eu II &     & ?  $\odot$ \\
4130.88 & Si II & M 2 & 2-3  $\odot$ \\
4136.92 & Mn II &  & about $\odot$  \\
4205.04 & Eu II &  &  70  $\odot$   \\
4215.52 & Sr II &  &   1 $\odot$    \\
4305.44 & Sr II &  &   1 $\odot$    \\
4558.65 & Cr II & M 1  &  10  $\odot$      \\
\hline
\end{tabular}
\caption{Classification lines and determined abundances for HD 67044}  
\end{table}  
%------------------------------------------------

\section{Model atmospheres and spectrum synthesis }

The effective temperature and surface gravity of HD 67044 were first evaluated using $B-V=-0.04$ and the effective temperature calibration versus $B-V$ in \cite{under82}. This yields $T_{eff} = 10200 $ K in good agreement with \cite{Huang2010} who derived $T_{eff} = 10519$K and $\log g = 3.72$ which we have adopted for the surface gravity.

A plane parallel model atmosphere assuming radiative equilibrium and hydrostatic equilibrium has been first computed using the ATLAS9 code \citep{Kurucz92}, specifically the linux version using the new ODFs maintained by F. Castelli on her website. The linelist was built starting from Kurucz's (1992) gfhyperall.dat file\footnote{http://kurucz.harvard.edu/linelists/} which includes hyperfine splitting levels.
This first linelist was then upgraded using the NIST Atomic Spectra Database\footnote{http://physics.nist.gov/cgi-bin/AtData/linesform} and the VALD\footnote{http://vald.astro.uu.se/~vald/php/vald.php} database operated at Uppsala University \citep{kupka2000}.

%------------------------------------------------

%------------------------------------------------

\section{Evidence for Si, Ti, Cr, Y, Zr and Eu excesses}

A grid of synthetic spectra was computed with SYNSPEC48 \citep{Hubeny92} to model the Si II, Ti II, Cr II, Mn II, Fe II, Y II, Zr II and Eu II lines. Computations were iterated varying the unknown abundance $[\frac{X}{H}]$ until minimisation of the chi-square between the observed and synthetic spectrum. Figure 1 displays the synthesis of 3 lines: Y II 3982.44 \AA, Zr II 3990.96 \AA, and Zr II 3998.82 \AA. The observed line profiles, rectified to the red wing of $H_{\epsilon}$ are compared to the synthetic spectrum providing the best fit to the oberved one. The model is computed for an overabundance of Yttrium of 200 $\odot$
and a Zirconium overabundance of 70 $\odot$  (solid line: observed normalised spectrum, dashed lines: synthetic spectrum).

% figure Y II

\begin{figure}[ht!]
 \centering
  \includegraphics[width=0.8\textwidth,clip]{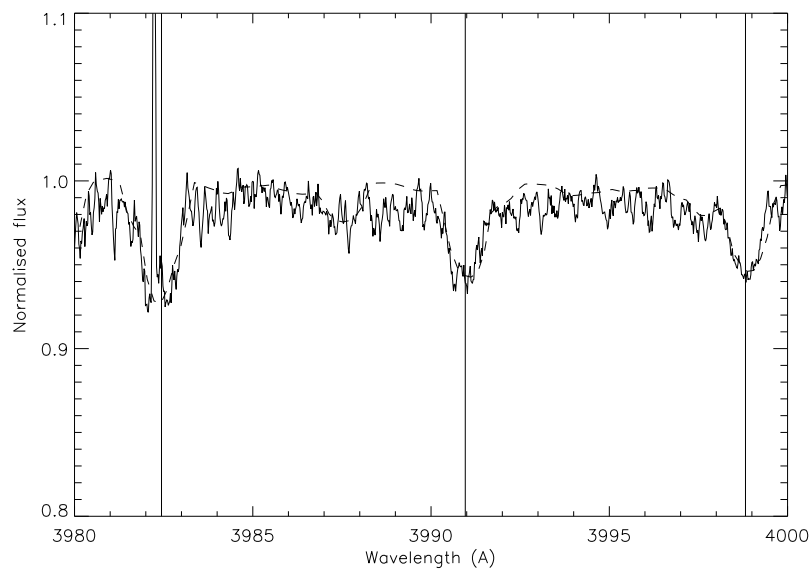}  \vskip -8.5cm   
  %% Note the ABSENCE of the extension .pdf , .eps or .ps  !
   \caption{Synthesis of the Y II and Zr II lines in the red wing of $H_{\epsilon}$ (observed: solid line, model: dashed line)}
   \label{fig2}
\end{figure}

\section{Conclusions}

% move to conclusion following text
HD 67044 has overabundances  which are characteristic of an SiCrEu star. We thus propose that {it should be reclassified as a late Chemically Peculiar B star of the SiCrEu type. It displays a mild overabundance of Silicon and large overabundances of Titanium, Chromium, Yttrium, Zirconium and Europium which range from 10 $\odot$ to about 200 $\odot$.
We are currently determining more abundances for this peculiar star.

% Optional acknowledgements
% -------------------------
\begin{acknowledgements}
The authors acknowledge very efficient support from the night assistants at Observatoire de Haute Provence. They have used the NIST Atomic Spectra Database and the VALD database operated at Uppsala University (Kupka et al., 2000) to upgrade atomic data.
\end{acknowledgements}

%%-----------------------------
%%   Bibliography
%%-----------------------------
%%
%% The reference list should contain all the references cited in the text, ordered alphabetically by surname (with
%% initials following). If there are several references to the same first author, they should be entered according
%% to the following scheme:
%% 1. One author: chronologically
%% 2. Author, one co-author: alphabetically by co-author, then chronologically
%% 3. Author, two or more co-authors: chronologically.
%%
%% Please note that for papers that have more than five authors, only the first three should be given, followed
%% by "et al."
%%
%% The format for references is the one adopted by A&A (see the example below).
%%
%% To set the reference list in the proper A&A format, we encourage you to use BibTEX and the natbib
%% package instead of the standard 'thebibliography' environment.
%%

%% The following lines are required when using BibTEX (strongly encouraged!):
\bibliographystyle{aa}  % A&A bibliography style file (aa.bst)
\bibliography{sf2a-template} % your references in file: Yourfile.bib

\end{document}